\documentclass[jkps,fleqn]{revtex4}
\usepackage[pdftex]{graphicx}
\usepackage{subcaption}
\usepackage{amssymb}
\usepackage{amsmath}
\usepackage{bm}
\usepackage{enumitem}
\usepackage{etoolbox}
\usepackage{caption}
\usepackage{lineno}

\begin{document}
\setcounter{page}{1}
\title[]{Towards the phenomenological implications of the \\ Physical Scheme in PDF fits\footnote{Presented at DIS2022: XXIX International Workshop on Deep-Inelastic Scattering and Related Subjects, Santiago de Compostela, Spain, May 2-6 2022. Speaker: C. Flett.}}
\author{Valerio Bertone$^1$}
\author{Chris A. Flett$^{2,3}$}
\author{Alan D. Martin$^4$}
\author{Misha G. Ryskin}
\address{$^1 $IRFU, CEA, Universit\'e Paris-Saclay, F-91191 Gif-sur-Yvette}
\address{$^2 $University of Jyv\"{a}skyl\"{a}, Department of Physics, P.O. Box 35, FI-40014 University of Jyv\"{a}skyl\"{a}, Finland}
\address{$^3 $Helsinki Institute of Physics, P.O. Box 64, FI-00014 University of Helsinki, Finland}
\address{$^4 $Institute for Particle Physics Phenomenology, Durham University, Durham, DH1 3LE, U.K.}


\begin{abstract}
We describe the impact on PDF extractions of the ‘Physical Scheme’, a heavy-quark mass scheme that accounts for the effects of intrinsic heavy-quarks and provides a way to smoothly transition over the 
heavy-quark thresholds.  The modifications made to the conventional $\overline{\text{MS}}$ scheme splitting functions and $\alpha_s$ running will be emphasised, collectively giving rise to a new Physical Scheme PDF evolution. We also present the analytic formulae for the DIS coefficient functions at NLO in the Physical Scheme and show results for the $F_2$ charm DIS structure function. Motivated by the upcoming statistics from the High-Luminosity LHC, embodying the next phase of the precision physics era at the LHC, as well as in time from the Electron-Ion collider,
we end by highlighting the need to quantify the effect of these corrections on PDF extractions over a wide range of momentum fractions $x$ and scales $Q^2$ and propose the use of $\texttt{xFitter}$ for doing so.

\end{abstract}

\pacs{}

\keywords{}

\maketitle

\section{INTRODUCTION}

The Physical Scheme, or Heavy Quark Smooth Threshold (HQST) Scheme, is a general mass flavour number scheme originally devised quite a number of years ago now  in\cite{deOliveira:2013tya}\footnote{with some minor corrections more recently given in\cite{Martin:2019dpw}.} and accounts for the effects of intrinsic heavy quarks already at the level of the splitting functions and provides a way to smoothly transition over the 
heavy quark thresholds. It does this by explicitly retaining the heavy quark mass corrections $m_h^2/Q^2$ in the splitting functions, and including heavy quarks in the renormalisation of $\alpha_s$ (which leads to a modified $\alpha_s$ running) as well as new coefficient functions in this scheme. In these proceedings we will be concerned with the application of this scheme to the description of neutral current ($\gamma, Z$) initiated Deep-Inelastic-Scattering (DIS).
Here, $m_h$ is the mass of the heavy quark and $Q$ is the virtuality of the probe in the typical DIS set up. The upshot of this is that, as will be discussed, the scheme resolves the kinks we see in the evolution and resolves discontinuities around threshold regions $Q^2$ around $m_h^2$.

Typically the global PDF analyses are performed to next-to-next-to-leading-order (NNLO) but in the first instance we shall study the effect of this scheme on PDF extractions at next-to-leading order (NLO), where it was shown\cite{deOliveira:2013tya, Martin:2019dpw} that to NLO accuracy, the effects of these power mass corrections need to be incorporated only into the conventional leading-order (LO) $\overline{\text{MS}}$ splitting functions and $\alpha_s$ running.  

To this end, we find it useful to ascertain the impact of this scheme on PDF fits at NLO using the public fitting tool, $\texttt{xFitter}$\cite{xFitterDevelopersTeam:2017xal}, and quantify the precision over a wide range of $x$ and $Q^2$ kinematics, relevant for current LHC measurements and in particular for the upcoming statistics from the High Luminosity LHC (HL-LHC) and the Electron-Ion collider~(EIC). We foresee a roll out of the Physical Scheme into this tool as part of the available heavy quark mass schemes, allowing users in time to perform their own phenomenological studies.



\section{Evolution of $\alpha_s$  }

\begin{figure}[t!]
\includegraphics[width=10cm]{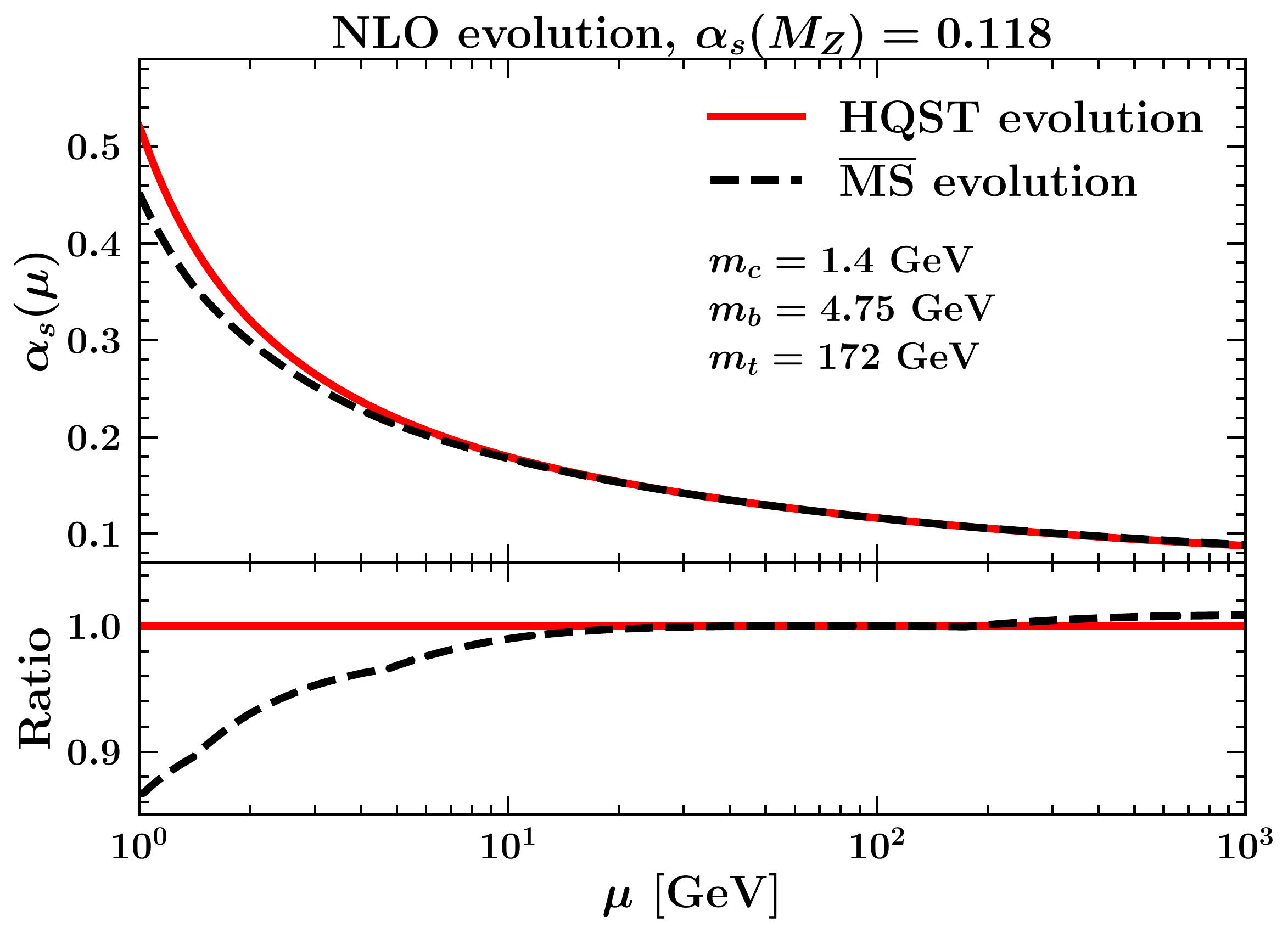}
\caption{{\bf Upper panel}: $\alpha_s$ running at NLO according to the HQST (solid, red curve) and $\overline{\text{MS}}$ (dashed, black curve) scheme. {\bf Lower panel}: ratio of the two evolutions, in which cusps displayed by the $\overline{\text{MS}}$ evolution are apparent in the transition over the heavy quark thresholds.}  \label{alphas}
\end{figure}

In Fig.~\ref{alphas}, we show the running of the strong coupling constant, $\alpha_s$, over a large range of scales at NLO. The dashed, black curve gives the $\alpha_s$ evolution according to the $\overline{\text{MS}}$ scheme, used for example in the global PDF analysis\cite{Bailey:2020ooq},  while the solid, red curve gives the $\alpha_s$ evolution according to the Physical Scheme.  The enhancement due to the heavy quark in the gluon propagator leads to a modified $n_f$ term in the beta function, $\kappa(\xi)$, which admits the following functional form
\begin{equation}
    \kappa(\xi) = 1 - 6 \xi + 12 \frac{\xi^2}{\sqrt{1+4\xi}} \ln \frac{\sqrt{1+4\xi}+1}{\sqrt{1+4\xi}-1},
\end{equation} 
where $\xi = m_h^2/\mu^2$. As $\mu^2$ gets large, $\kappa(\xi) \rightarrow 1$ and at moderate values of the scale it is a positive definite function between zero and one. Since $\kappa(\xi) < 1$ and the terms proportional to $n_f$ in $\beta_0$ and $\beta_1$ are always positive, it follows that $\beta_0^{\text{HQST}} < \beta_0^{\overline{\text{MS}}}$ i.e. that the $\overline{\text{MS}}$ running is flatter than the HQST one. Since at $M_Z$ the two must be equal due to the boundary condition $\alpha_s(M_Z) = 0.118$, it follows that we have the hierarchy 
\begin{equation}
   \alpha_s^{\text{HQST}}(\mu) > \alpha_s^{\overline{\text{MS}}}(\mu),\,\,\,\,\,\text{for}\,\,\,\,\, \mu < M_Z \,\,\,\,\,\,\,\,\,\,\text{and}\,\,\,\,\,\,\,\,\,\,
    \alpha_s^{\text{HQST}}(\mu) < \alpha_s^{\overline{\text{MS}}}(\mu),\,\,\,\,\,\text{for}\,\,\,\,\, \mu > M_Z. 
    \end{equation}
This leads to a consistent enhancement of the $\alpha_s$ running from low scales up to the $Z$ mass, as the red curve shows. In the lower panel of Fig.~\ref{alphas}, we show the ratio of the two evolutions in which the kinks displayed by the $\overline{\text{MS}}$ evolution at the heavy quark thresholds at NLO are clearly displayed. These kinks arise because the $\overline{\text{MS}}$ evolution is discontinuous at threshold, due to the step function from a given $n_f$ to $n_f + 1$.
This is avoided in the Physical Scheme by having the heavy quark mass dependence explicit rather than being activated at certain thresholds. Consequently, the HQST evolution is smooth and there are no kinks.

In the Physical Scheme to NLO, the heavy quarks are encoded intrinsically within explicit power mass corrections embodied within modified splitting functions at LO as follows,
\begin{align}
    P_{hg} (\xi,x) &= 2\, T_R\, \eta \left( x^2+(1-x)^2+(1-\eta) 2x(1-x) \right) \theta(\eta-4x+3\eta x)\,, \nonumber \\
    P_{hh}^{\rm real}(\xi,x) &= 2\,C_F\,\left( \frac{1+x^2}{1-x}\frac{1}{1+\xi(1-x)}+x(x-3)\frac{\xi}{(1+\xi(1-x))^2}\right), \nonumber \\
    P_{gh}(\xi,x) &= 2\,C_F\,\left(\frac{1+(1-x)^2}{x}\frac{1}{1+\xi x}+(x-1)(x+2)\frac{\xi}{(1+\xi x)^2}\right)\,, \nonumber \\
P_{gg}(\xi,x) &=P_{gg}^{(n_l)}(x)-\delta(1-x)2\,T_R\,\sum_h\frac{\eta_h^3}{(4-3\eta_h)^2}\left(1-\frac43\frac{\eta_h^2}{(4-3\eta_h)}+\frac{\eta_h^3}{(4-3\eta_h)^2}\right)\,,
    \label{hqstsf}
\end{align}
where $\eta = 1/(1+\xi)$ and $P_{gg}^{(n_l)}$ is the $\overline{\rm MS}$ result with $n_l$ active flavours and $h$ runs only over
the massive quarks in the last equation. It is clear that $P_{hh}^{\rm real}$ and $P_{gh}$ are related by a $x \rightarrow 1-x$ symmetry and the inclusion of the heavy quark mass is manifest within the parameter $\xi$. As $\mu^2 \rightarrow \infty$ ($m_h \rightarrow 0$), the parameter $\xi \rightarrow 0$ and these reduce back to the familiar conventional $\overline{\text{MS}}$ distributions.

To emphasise, there is no matching procedure employed in the Physical Scheme, the heavy quarks are active at all scales and encapsulated in a smooth fashion through the scale and mass dependent $\xi$ factors in the splitting functions, as well as through a modified $\alpha_s$ running. 


\section{Physical Scheme coefficient functions}

\begin{figure}[t!]
\includegraphics[width=10cm]{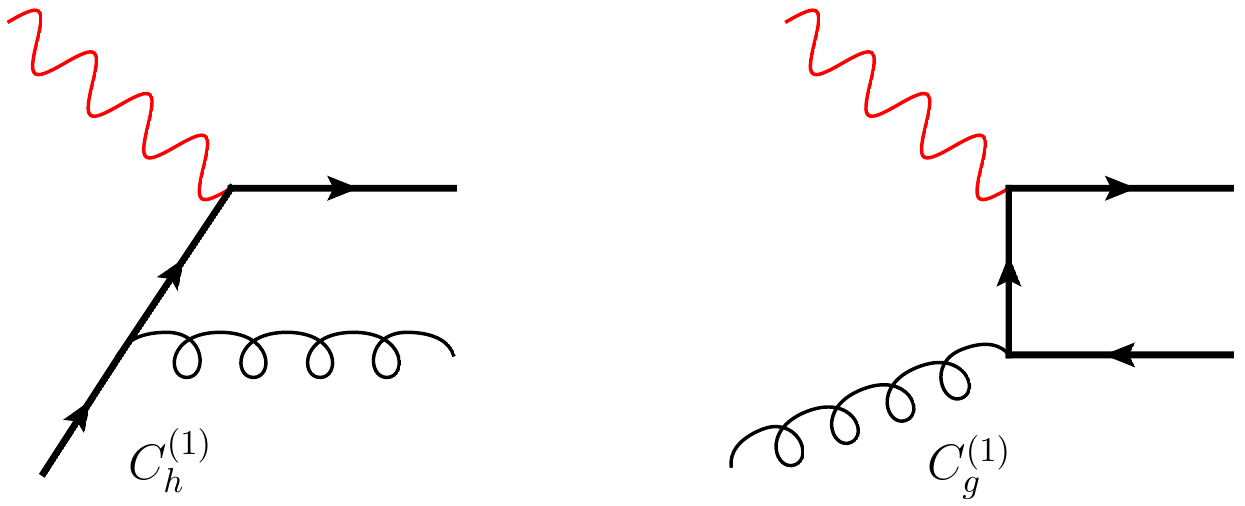}
\caption{NLO Feynman diagrams for heavy quark (left) and gluon (right) initiated DIS. Bold lines represent heavy quarks.}  \label{NLOdiags}
\end{figure}

We now turn to discuss the DIS coefficient functions in the Physical Scheme to NLO, for both heavy quark and gluon initiated processes\footnote{Here written for photon (neutral current) induced DIS, but the corresponding formulae are deduced analogously for neutral current $Z$.} 
$\gamma^* h \rightarrow h g$ and $\gamma^* g \rightarrow h \bar{h}$, see Fig.~\ref{NLOdiags}, where $h$ labels a heavy quark, $\bar{h}$ a heavy antiquark and $g$ a gluon. In the same spirit as the ACOT scheme\cite{Aivazis:1993kh, Aivazis:1993pi}, the Physical Scheme DIS coefficient functions to NLO are constructed as follows,
\begin{equation}\label{eq:ACOT}
C_a^{(1),\rm HQST}(x,Q,m_h) = C_a^{(1),\rm FFNS}(x,Q,m_h) -
\frac{\alpha_s(Q)}{4\pi}\sum_{b=g,q,h}
C_b^{(0)}(x,Q,m_h) \otimes\int_{0}^{Q^2}\frac{d\mu^2}{\mu^2} \left(P_{ba}^{(0),\rm
    HQST}(\xi,x) - P_{ba}^{(0),\rm
    \overline{\text{MS}}}(x)\right) \,,
\end{equation}
where $a = h,g$ and $q$, a light quark.  In all cases we have the Fixed-Flavour-Number-Scheme (FFNS) coefficient function\footnote{See\cite{Forte:2010ta} and\cite{Kretzer:1998ju} for $C_g$ and $C_h$, respectively, to NLO.} and, in the subtraction term that avoids a double counting and removes the logarithmic divergence of the function at large scales, we replace the $\overline{\text{MS}}$ splitting functions by those which describe the evolution in the HQST scheme. 
Here, $P^{(0),\text{HQST}}_{ba}$ are the modified LO splitting functions given in eqn.~\eqref{hqstsf} and $C_b^{(0)}$ are the LO Physical scheme coefficient functions, 
\begin{align}
C_q^{(0)}(x) &= e_q^2\delta(1-x), \nonumber \\
C_h^{(0)}(\xi,x) &= e_h^2\zeta(2-\zeta)\delta\left(1-\frac{x}{\zeta}\right), \\
C_g^{(0)}(x) &= 0,  \nonumber
\end{align}
with $\zeta = 2/(1+\sqrt{1+4\xi})$.  Note that the HQST scheme splitting functions are used for the subtraction term, and the fact that the LO coefficients of the expansion of the splitting functions in the HQST scheme
depend on the factorisation scale $\mu$ implies an explicit integral over $\mu$ from zero up to the virtuality $Q^2$. Note also that the ACOT scheme is typically implemented in its simplified form (the S-ACOT scheme). In this variant the heavy quark mass $m_h$ is set to zero in the computation of diagrams with incoming heavy quarks and/or internal on-shell cuts on a heavy-quark line. In
other words, in the S-ACOT scheme the heavy-quark initiated diagrams are never included and the subtraction terms are always massless.

The explicit integration over $\mu$ of the HQST splitting functions generates the correct $\theta$ function so that the kinematic support of each of these two contributions is the same. For the gluon contribution, it is the $\theta$ function which enforces the on-shell production of two heavy quarks. To emphasise, the FFNS coefficient functions are valid for $Q \sim m_h$, but the Physical Scheme is applicable for all scales. As the FFNS terms may exhibit logarithmic divergences in the ratio $Q^2/m_h^2$, there is a need to remove these by performing the (massive) convolution $\sim C_a^{(0)} \otimes P_{ba}$ in the subtraction term so that the Physical Scheme coefficient functions are well defined for large scales, too.

\section{PDF evolution and $F_2^{\lowercase{c}}$ structure function}

Having discussed the HQST splitting functions and NLO DIS coefficient functions, we are in a position to consider the evolutions of quark and gluon PDFs in this scheme. To do this we take the NLO MSHT20 parton set\cite{Bailey:2020ooq} and transform them to the Physical Scheme using the transformation
\begin{equation}
a^{\rm{HQST}}(x,\mu^2)=a^{\overline{\text{MS}}}(x,\mu^2) + \frac{\alpha_s}{4\pi} \int dz\sum_{b}\delta_{ab}(z) b^{\overline{\text{MS}}}(x/z,\mu^2)\,,
\label{sch}
\end{equation}
see eqn.~(6) in~\cite{deOliveira:2013tya}, at the initial scale $Q_0 = 1.5$ GeV to convert the MSHT20 $\overline{\text{MS}}$ distribution to a HQST one before evolution. Explicitly, the HQST distribution at initial scale $Q_0$ is calculated via eqn.~\eqref{sch} and then evolved up to the final scale $\mu$ in the HQST scheme.

 \begin{figure*}
        \centering
        \begin{subfigure}[b]{0.32\textwidth}
            \centering
            \includegraphics[width=\textwidth]{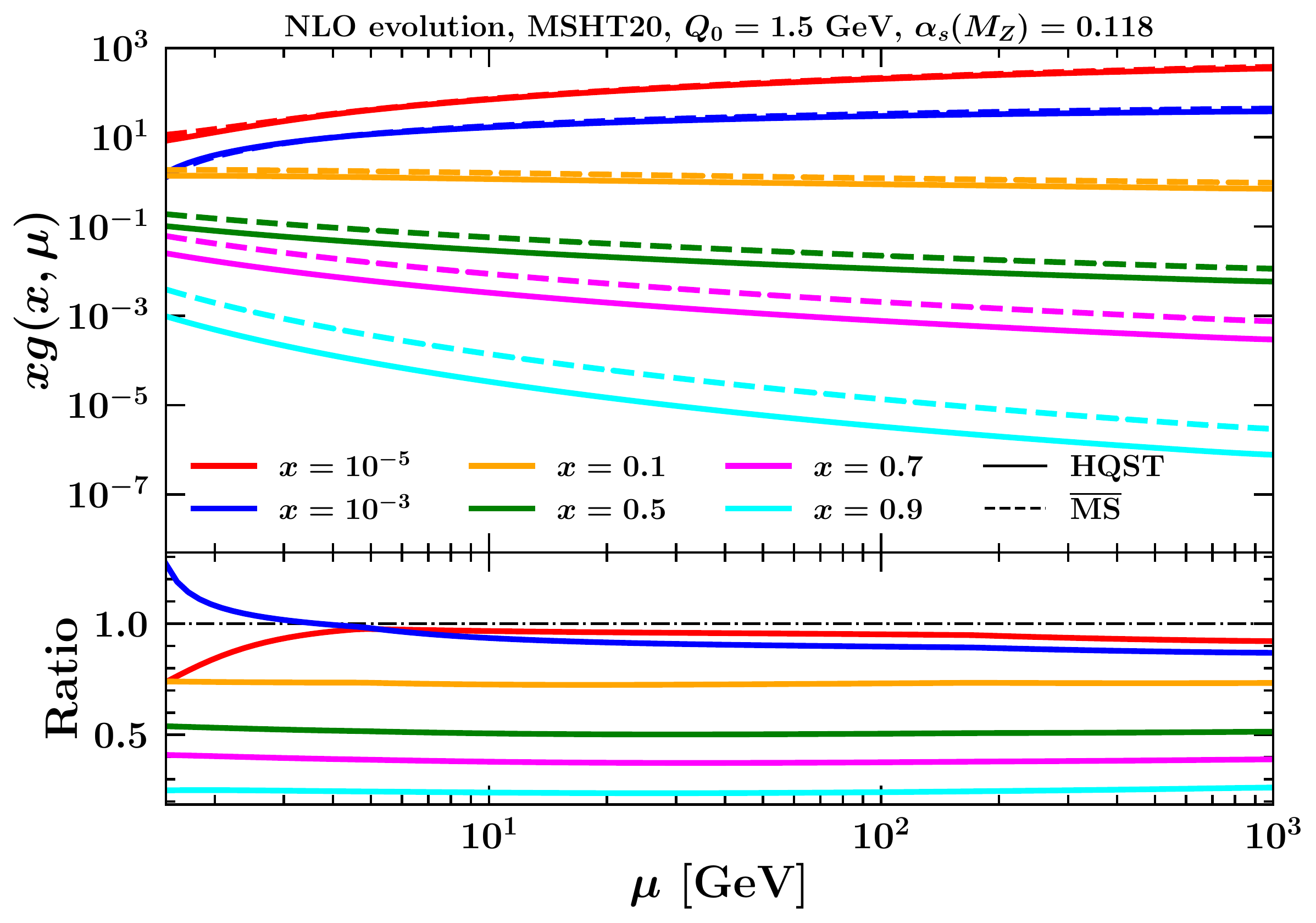}
            {{\small}}    
            \label{fig:jpsi5decomps}
        \end{subfigure}
        \begin{subfigure}[b]{0.32\textwidth}  
             \centering
            \includegraphics[width=\textwidth]{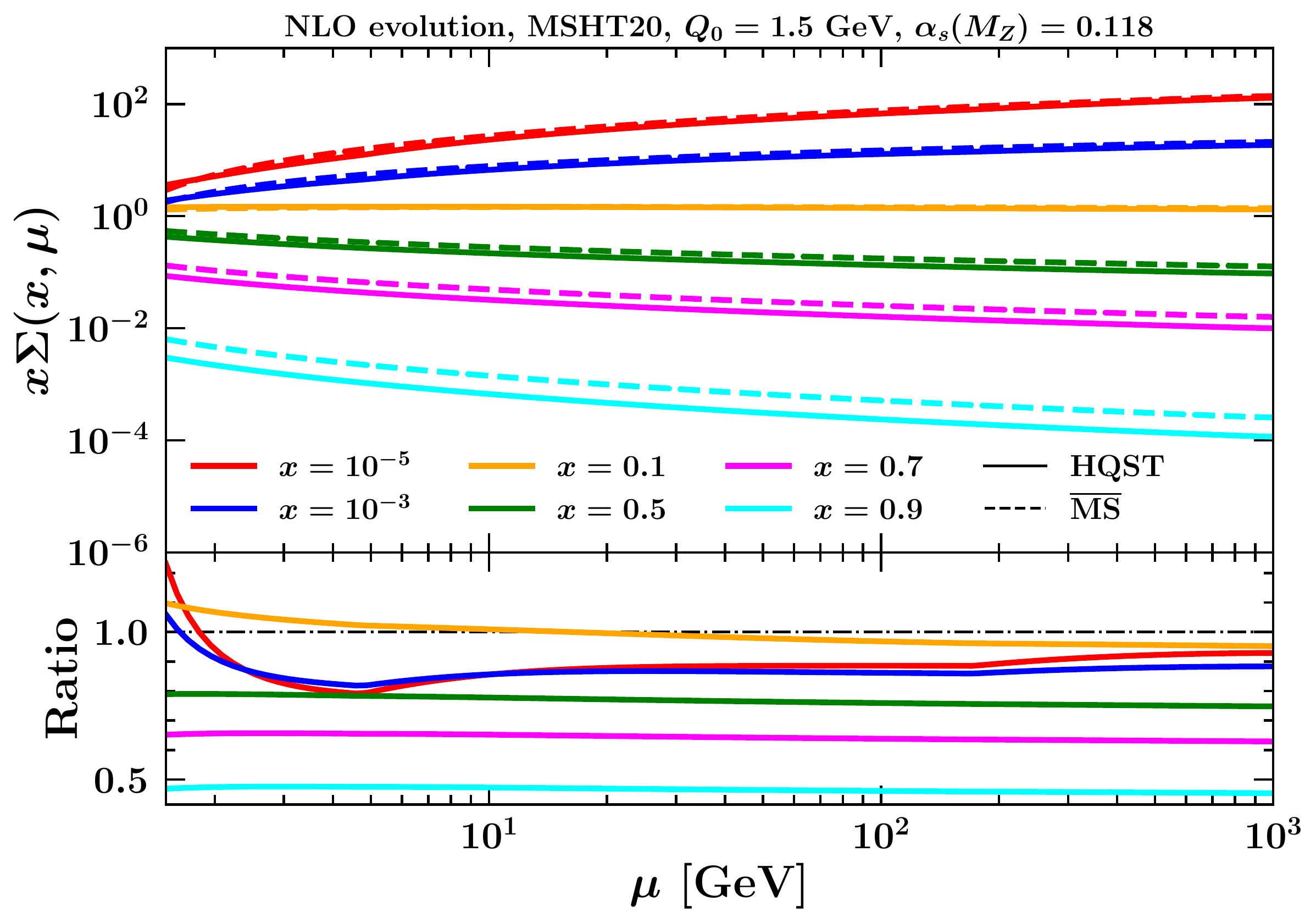}
            {{\small }}    
            \label{fig:jpsi5data}
        \end{subfigure}
        \begin{subfigure}[b]{0.32\textwidth}   
             \centering
            \includegraphics[width=\textwidth]{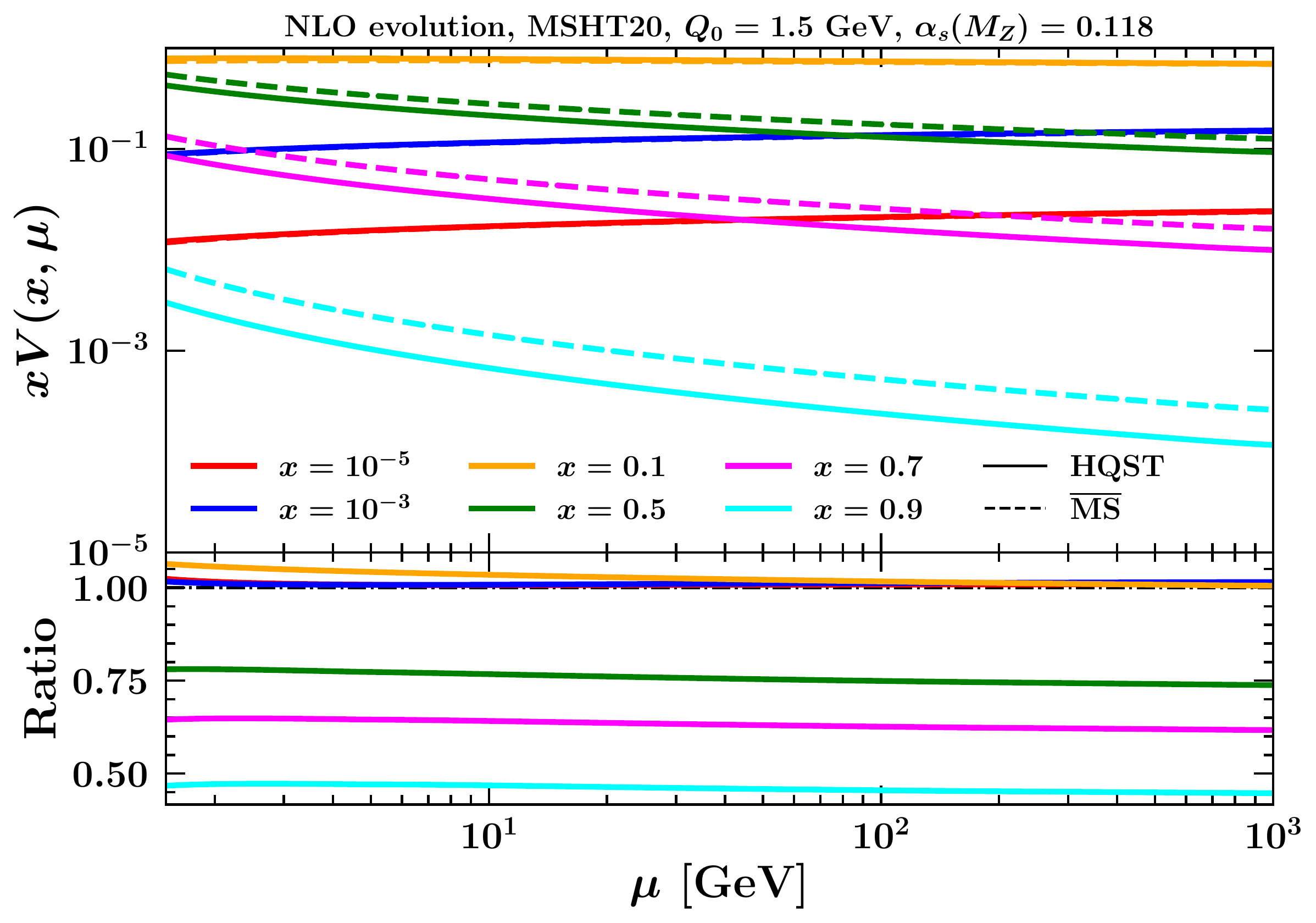}
            {{\small }}    
            \label{fig:ups5decomps}
        \end{subfigure}
        \vskip\baselineskip
        \begin{subfigure}[b]{0.32\textwidth}   
             \centering
            \includegraphics[width=\textwidth]{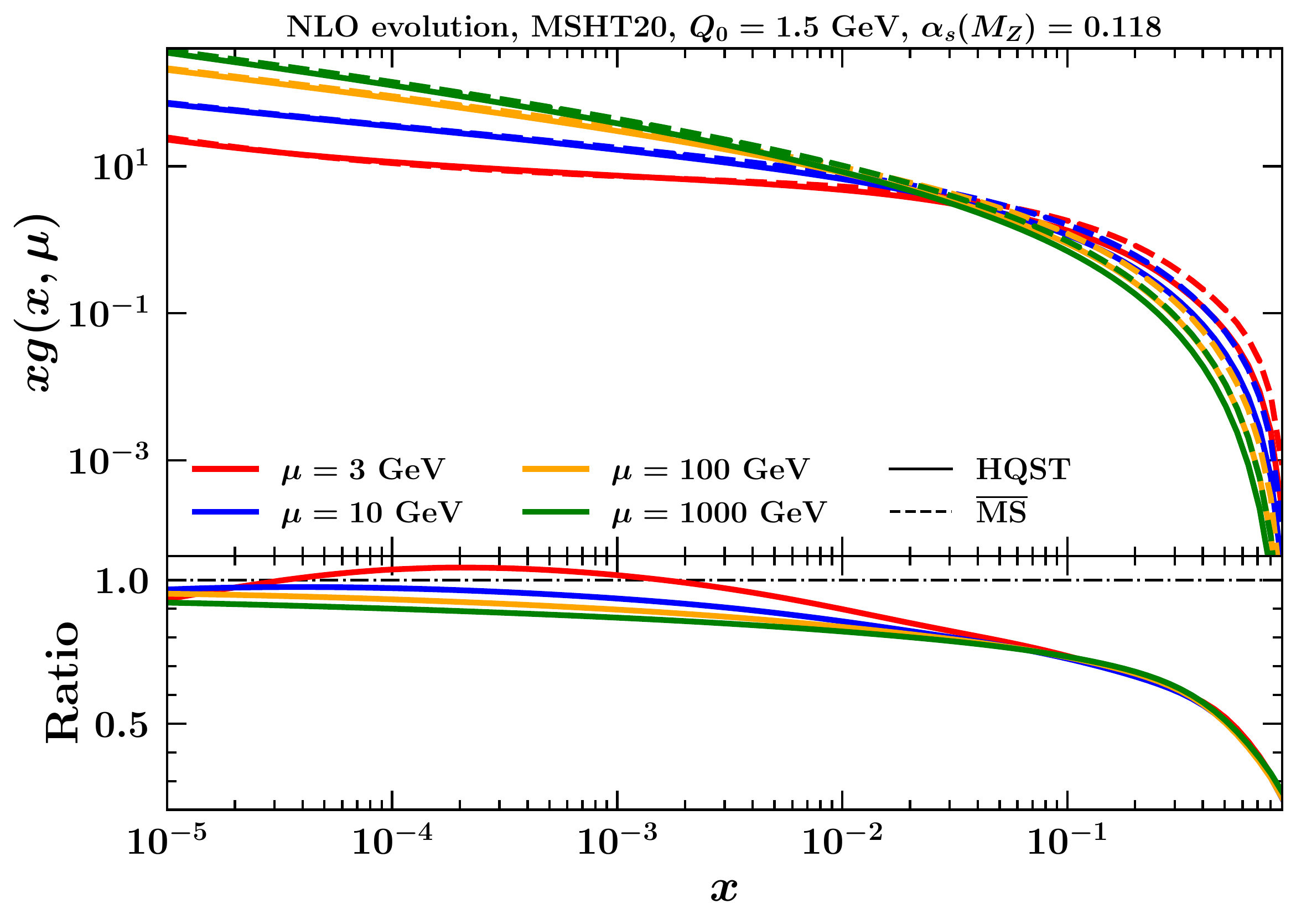}
            {{\small }}    
            \label{fig:ups5data}
        \end{subfigure}
        \begin{subfigure}[b]{0.32\textwidth}   
             \centering
            \includegraphics[width=\textwidth]{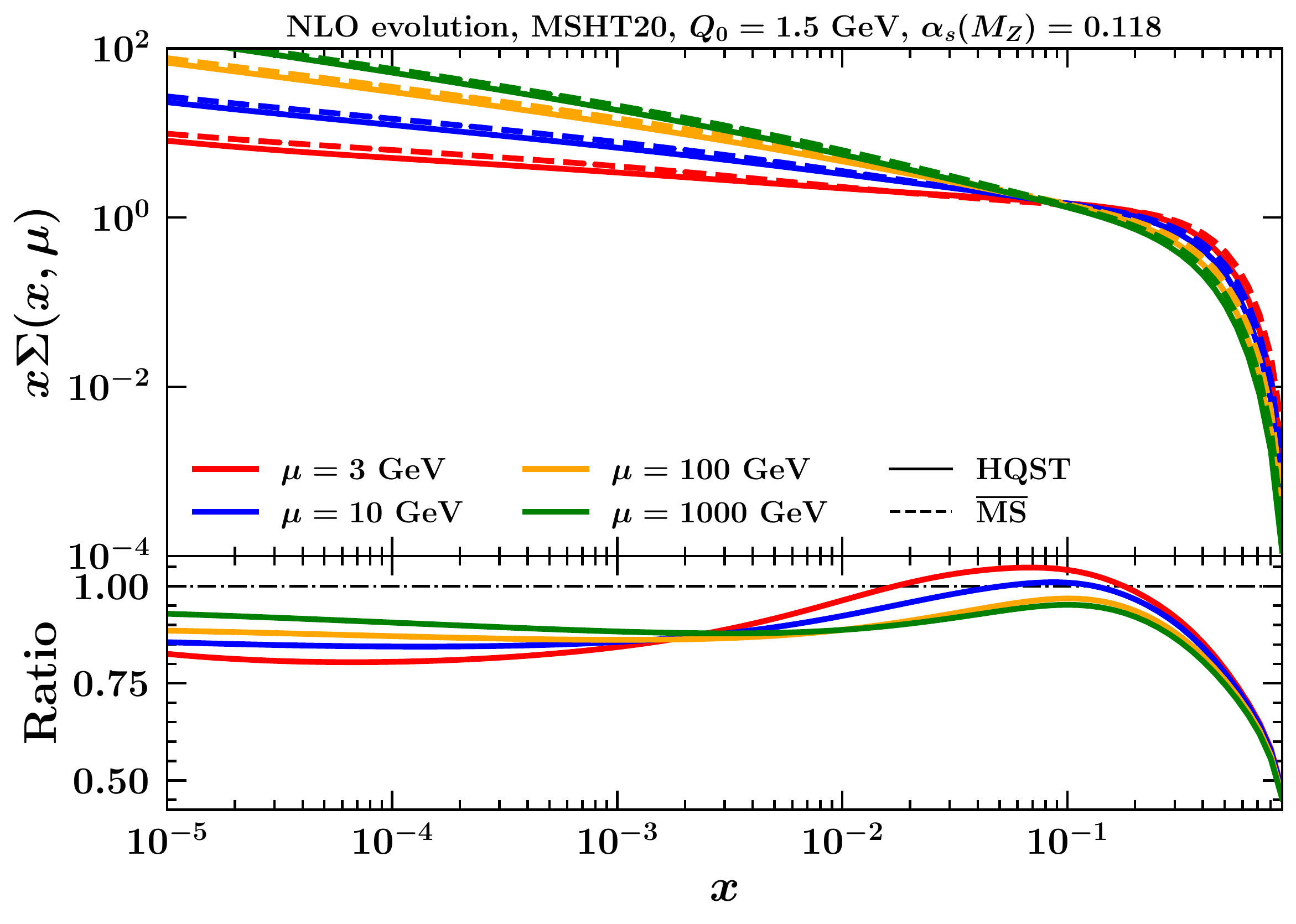}
            {{\small }}    
            \label{fig:ups5data}
        \end{subfigure}
        \begin{subfigure}[b]{0.32\textwidth}   
             \centering
            \includegraphics[width=\textwidth]{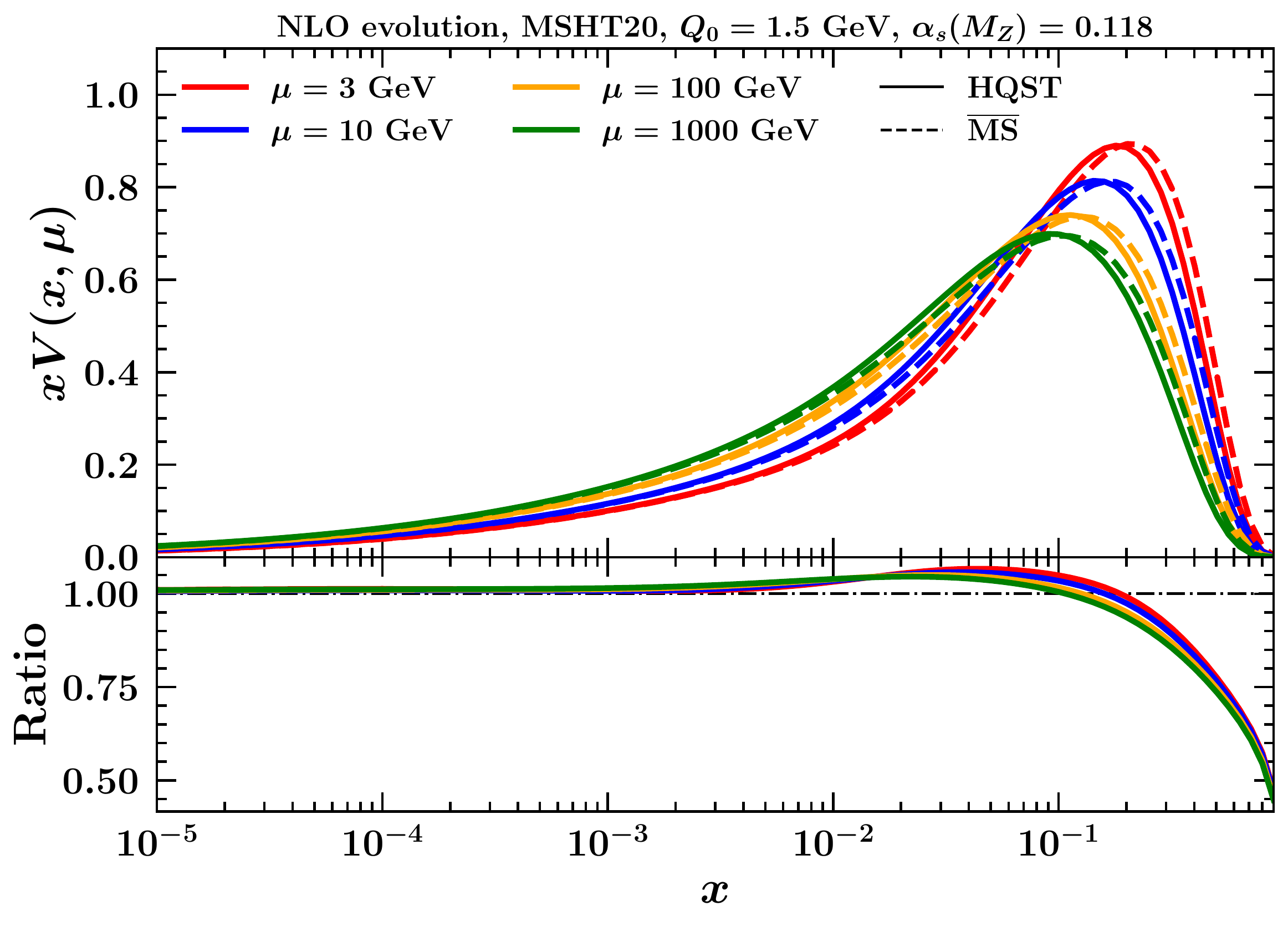}
            {{\small }}    
            \label{fig:ups5data}
        \end{subfigure}
        \caption[  ]
        {Comparison of PDF evolution at NLO in the HQST (solid) and $\overline{\text{MS}}$ (dashed) schemes. {\bf Upper row}: evolution as a function of $\mu$ for fixed $x$. {\bf Lower row}: evolution as a function of $x$ for fixed $\mu$.} 
        \label{fig:evol}
    \end{figure*}

\begin{figure}[t!]
\includegraphics[width=10cm]{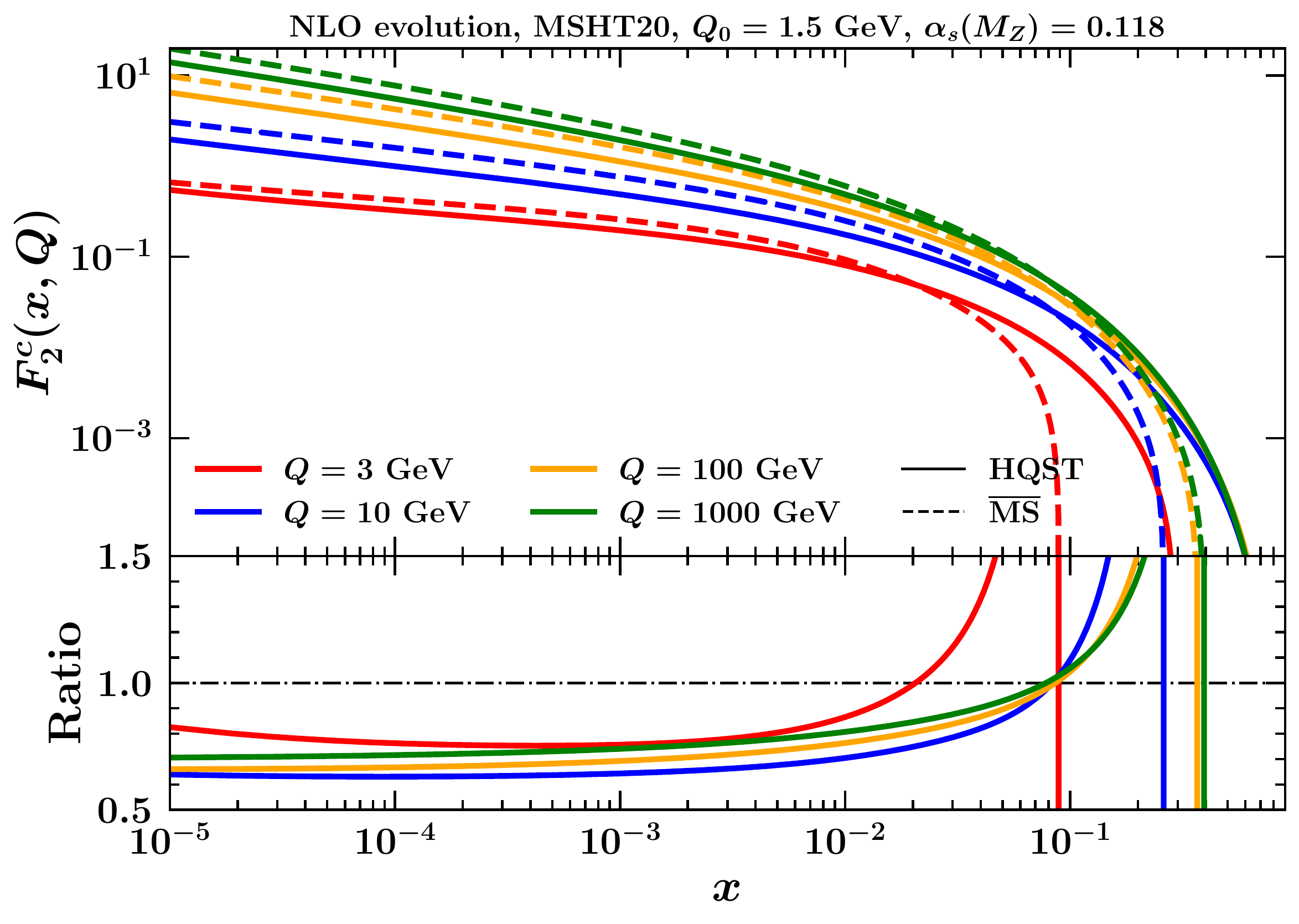}
\caption{Comparison of the DIS charm structure function, $F_2^c$, as a function of $x$ for  fixed $Q^2$. The solid (dashed) curves are predictions in the HQST $(\overline{\text{MS}})$ scheme.}  \label{fig:F2c}
\end{figure}

In Fig.~\ref{fig:evol} we compare the parton distributions and in Fig.~\ref{fig:F2c} we use them to compare the resulting behaviour of the $F_2$ charm structure functions. The results show that the effect of going through the heavy quark thresholds is smooth.  However, the size of the changes in going from $\overline{\text{MS}}$ to the Physical Scheme is unrealistic.  We are studying the cause of this effect in detail.

The best way to obtain realistic smooth PDFs is to perform a completely new global analysis working from the beginning in the Physical Scheme and fitting not only with Physical Scheme input, but with Physical Scheme $\alpha_s$ as well.  A comparison with $\overline{\text{MS}}$ partons fitted to the same data can then be made.

\newpage
\section{Conclusions}

In these proceedings, we have described the Physical Scheme or Heavy Quark Smooth Threshold Scheme and its application to DIS at NLO. This scheme retains explicit heavy quark mass corrections through modified splitting functions and $\alpha_s$ running, resolving the kinks in $\alpha_s$ in the $\overline{\text{MS}}$ scheme and providing for a smooth transition over the 
heavy quark thresholds. 

The difference in the PDF evolution at NLO between the $\overline{\text{MS}}$ and the Physical Scheme shows that further study is needed in order to make a realistic comparison, and hence to demonstrate the effect of the smooth behaviour through the heavy quark thresholds in the Physical Scheme; see the last paragraph of Section IV.


Motivated by the upcoming statistics from e.g. the HL-LHC within the era of precision physics measurements of the LHC programme, as well as at the EIC, we find it an important endeavour to quantify the impact of these heavy quark mass corrections on PDF fits at a given fixed order. The state of the art of global PDF analyses is now NNLO, but the size of the effects will first be quantified at NLO. To accomplish this, the public PDF fitting tool $\texttt{xFitter}$ will be employed to fit PDFs to DIS data over a wide range of momentum fractions $x$ and scales $Q^2$.


\end{document}